\documentclass[12pt]{iopart}

\usepackage{ae,aecompl}
\usepackage[T1]{fontenc}
\usepackage[latin1]{inputenc}
\usepackage{verbatim}
\usepackage{textcomp}
\usepackage{graphicx}
\usepackage{multirow}
\usepackage{subfigure}
\usepackage{color}{}
\usepackage{latexsym}
\definecolor{Red}{rgb}{1,0.0,0.0}

\begin{document}

\title{Effects of time reversal symmetry in dynamical decoupling}

\author{Alexandre M. Souza, Gonzalo A. \'{A}lvarez and Dieter Suter}
\address{Fakult\"at Physik, Technische Universit\"at Dortmund, D-44221,
Dortmund, Germany}


\begin{abstract}

Dynamical decoupling (DD) is a technique for preserving the coherence of quantum mechanical states
in the presence of a noisy environment. 
It uses sequences of inversion pulses to suppress the environmental perturbations by 
periodically refocusing them. 
It has been shown that different sequences of inversion pulses show vastly different
performance, in particular also concerning the correction of experimental pulse imperfections.
Here, we investigate specifically the role of time-reversal symmetry in the building-blocks 
of the pulse sequence.
We show that using time symmetric building blocks often improves the performance 
of the sequence compared to sequences formed by time asymmetric building blocks. 
Using quantum state tomography of the echoes generated by the sequences,
we analyze the mechanisms that lead to loss of fidelity and show how they can be
compensated by suitable concatenation of symmetry-related blocks of decoupling pulses.

\end{abstract}

\maketitle

\section{Introduction \label{sec1} }

Dynamical decoupling (DD) \cite{viola,yang} is becoming an established technique for preserving quantum states from 
decoherence with possible applications in quantum 
information \cite{ddgate1,ddgate2,ddgate3,ddgate4,dqc1dd} and 
magnetic resonance \cite{noise1,noise2,noise3,magnetometer1,meriles,magnetometer2,magnetometer3}. The 
technique was devised to increase decoherence times by refocusing the system-environment 
interactions with a sequence of control pulses periodically
applied to the system. 
Recent experiments have
successfully implemented DD methods and demonstrated
the resulting increase of coherence times in different 
systems  \cite{exp4,exp3,warren,pra,exp1,exp2,ashok,prl,jpb}. 
These works also showed that the performance of  
DD sequences can be limited or even counterproductive if the accumulated effect 
of pulse imperfections becomes too strong \cite{pra,exp2,prl,jpb}. 
One approach to compensate the effect of these errors is to combine
one basic decoupling cycle with a symmetry-related copy into a longer cycle.
The resulting cycle can be more robust, i.e. less susceptible to pulse imperfections
than its building blocks, provided the basic blocks are well chosen and combined in 
a suitable way.

In the field of nuclear magnetic resonance (NMR), symmetry-related arguments
have often been used for constructing supercycles \cite{Haeberlen1,mansfield,rhim1,rhim2}. 
Using the symmetry properties of specific interactions, it is possible to remove them selectively
while retaining or restoring others  \cite{levitt1,levitt2}. Symmetrization is widely used to eliminate
unwanted odd-order average Hamiltonian terms \cite{burum}. 
This approach has been instrumental in the design of high-performance decoupling and 
recoupling  sequences  \cite{levitt1,levitt2}.  
Besides sequence development, symmetry arguments have also 
been used extensively in the design of individual pulses with reduced sensitivity to 
experimental imperfections \cite{pines1,morris}.

The main goal of this paper is to investigate differences between otherwise identical DD cycles,
in which the timing of the pulses is either symmetric with respect to time reversal, or not.
As the basic block we consider the XY-4 sequence \cite{maudsley}. 
We compare the performance of the basic sequences as well as compensated higher-order sequences
and analyze their imperfections theoretically by average Hamiltonian theory \cite{Haeberlen1,ath} and experimentally 
by applying quantum state tomography \cite{chuang,oliveira} to 
the system after the end of each decoupling cycle.

This paper is organized 
as follows. 
In Section \ref{sec2} we introduce the basic idea of dynamic decoupling and demonstrate the relevance
of time reversal symmetry in this context.
In the subsequent  sections \ref{sec3} and \ref{sec4} we compare different sequences
based on symmetric or asymmetric building blocks.
In the last section we draw some conclusions.

\section{Symmetrization in DD \label{sec2}}

\begin{figure}[htbp]
\vspace*{13pt}
\begin{center}
{\includegraphics[width=13.0cm]{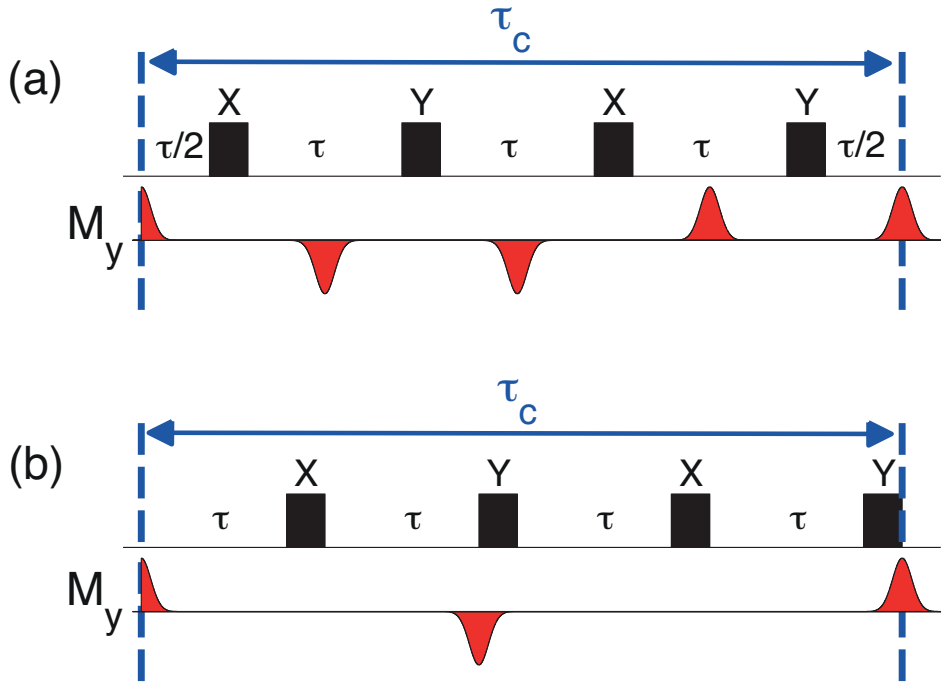}}
\end{center}
\vspace*{13pt}
\caption{\label{fig1} Schematic  representation of dynamical decoupling sequences: (a)  time symmetric XY-4 and (b) asymmetric XY-4.}
\end{figure}

Dynamical decoupling is a technique in which the coherence of qubits is dynamically 
protected by refocusing the qubit-environment interaction \cite{viola,yang}. Within this 
technique, a sequence 
of $\pi$ rotations 
is  periodically applied to the system. 
For a purely dephasing environment, i.e. one that couples only to the $z$-component of the system qubit,
this can be achieved simply by a train of identical $\pi$-pulses, the so-called 
Carr-Purcell (CP)- \cite{cp} or Carr-Purcell-Meiboom-Gill (CPMG)-sequence \cite{mg}.
The shortest DD sequence that
cancels the zero-order average Hamiltonian for a general system-
environment interaction \cite{viola,cdd} is the XY-4 sequence \cite{maudsley} (see figures 
\ref{fig1}a and \ref{fig1}b). This sequence also has the advantage of being much 
less sensitive to pulse imperfections than the CP-sequence \cite{maudsley,xy}. 

\begin{figure}[htbp]
\vspace*{13pt}
\begin{center}
\includegraphics[width=13.0cm]{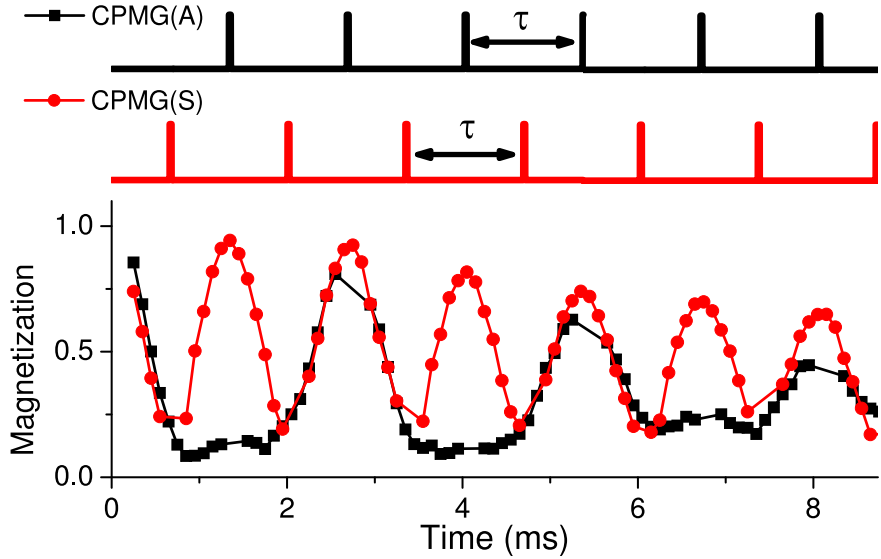} \\
\end{center}
\vspace*{13pt}
\caption{\label{cpmg1} Experimentally observed evolution of the $^{13}$C nuclear spin magnetization of adamantane during the symmetric
(red circles) and asymmetric (black squares) CPMG pulse sequences.}
\end{figure}

In the spectroscopy and quantum computing communities, two  versions of the XY-4
sequence are used that differ by a seemingly minor detail.
As shown in Fig.  \ref{fig1}a, the basic cycle originally 
introduced in NMR \cite{maudsley,xy} starts with a delay 
of duration $\tau/2$ and ends with another delay of the same duration.
It therefore shows reflection symmetry in the time domain with respect to the
center of the cycle. 
In contrast to that, the sequence used in the quantum information 
community \cite{viola,cdd,Khodjasteh,sdd,preskill}
starts with a delay of duration $\tau$ and ends after the fourth pulse (see Fig. \ref{fig1}b).
Clearly, this cycle is not symmetric in time.
One consequence of this small difference is that in the case shown in Fig. \ref{fig1}a,
the echoes are formed in the center of the windows between any two pulses,
while in Fig. \ref{fig1}b, the echoes coincide with every second pulse. 
The separation in time between the echoes is therefore twice as long in this case.

Figure \ref{cpmg1} illustrates this difference with an experimental example.
Here, we measured the time evolution of the $^{13}$C nuclear spin polarization during
a CPMG sequence, using in one case a time-symmetric and in the other case
a non time-symmetric cycle. 
The sample used for this experiment was polycrystalline adamantane.
The dephasing of the nuclear spin signal originates from the interaction with an environment
consisting of $^1$H nuclear spins.
To make this environment appear static and generate a long  train of echoes, 
we applied a homonuclear decoupling sequence to the protons \cite{ashok}.
As expected, in the symmetric case, the echoes appear with half the separation
of the asymmetric case.

\begin{figure}[htbp]
\vspace*{13pt}
\begin{center}
\includegraphics[width=9.0cm]{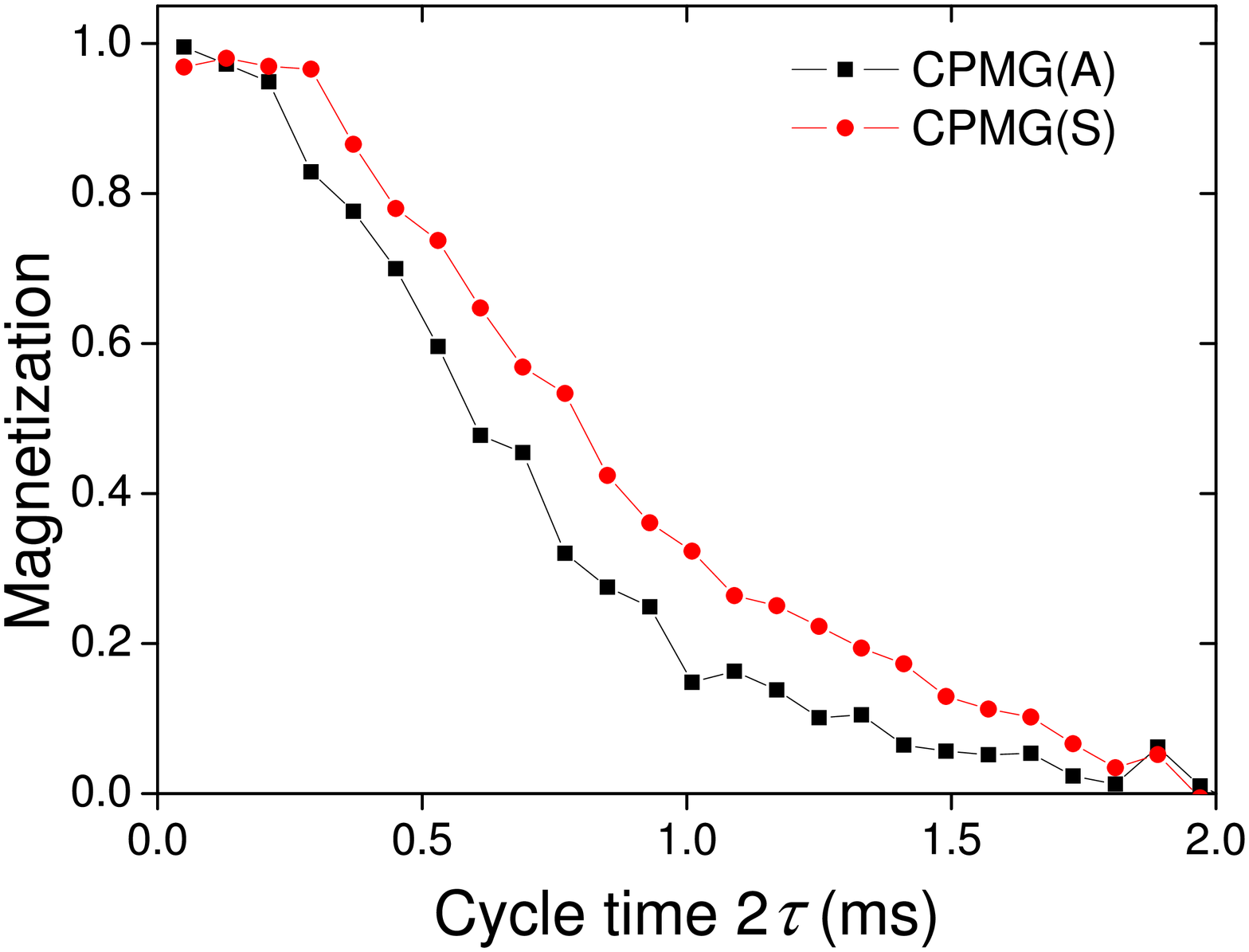} \\
\end{center}
\vspace*{13pt}
\caption{\label{cpmg} $^{13}$C nuclear spin magnetization after one cycle of symmetric 
and asymmetric CPMG sequences for different cycle times 2 $\tau$.}
\end{figure}

The larger separation of the echoes also can lead to a faster decay of the echo amplitude
if the environment is not static.
This is illustrated in Fig. \ref{cpmg}, which shows the decay of the echo amplitude as a function of time.
In this case, we did not apply homonuclear decoupling and the system-environment interaction
is therefore modulated by the homonuclear magnetic dipole-dipole interaction between the 
environmental protons \cite{ashok}.
The plotted signal shows the echo amplitude measured after a single CPMG cycle for the symmetric and asymmetric case,
as a function of the cycle time $2 \tau$.
Clearly the symmetric cycle is more efficient in preserving the state of the system -
in agreement with findings from multiple pulse NMR \cite{levitt1,levitt2}.

Since any multiple pulse cycle suffers from imperfections and non-ideal properties,
it is often desirable to construct longer cycles that have better properties than simply
repeating the basic cycle. 
Examples of DD sequences that can be constructed from the XY-4 cycle
include the  XY-8 and XY-16 \cite{xy} sequences shown in Table \ref{tab1}. 
Here, the XY-8 sequence concatenates an XY-4 cycle with its time-reversed version
\cite{xy,Khodjasteh,sdd,preskill},
thus generating a new cycle, which is inherently time-symmetric, independent of which
version of the XY-4 sequence was used for the building blocks.

In the following sections \ref{sec3} and \ref{sec4} we  show that two symmetric 
sequences, constructed according to the same rules from a basic XY-4  block can have 
different behaviors depending if the basic block is chosen to be symmetric or not. 

\begin{table}
\begin{centering}

{\footnotesize }\begin{tabular}{ccc}

 &  \textbf{\footnotesize Asymmetric}{\footnotesize{} } & \textbf{\footnotesize Symmetric}{\footnotesize{} }\tabularnewline
\hline 
 &  & \tabularnewline
{\footnotesize \bf{$XY-4$} } & {\footnotesize \bf{$[\tau-X-\tau-Y]^{2}$} } & {\footnotesize \bf{$[\tau/2-X-\tau-Y-\tau/2]^{2}$} }\tabularnewline
 &  & \tabularnewline
\hline 
 &  & \tabularnewline

{\footnotesize \bf{$XY-8$} } & \multicolumn{2}{c}{ {\footnotesize \bf{$(XY-4)(XY-4)^{T}$} } } \tabularnewline
 &  & \tabularnewline
{\footnotesize \bf{$XY-16$} } &  \multicolumn{2}{c}{ {\footnotesize \bf{$(XY-8)\overline{(XY-8)}$} } }\tabularnewline
 &  & \tabularnewline
\hline
\end{tabular}
\par\end{centering}{\footnotesize \par}

\caption{{\footnotesize \label{tab1} Dynamical decoupling sequences. 
The top line shows the time symmetric and asymmetric versions of XY-4, which can 
be used as building blocks for other sequences. 
$X$ and $Y$ represent $\pi$-pulses around the $x$
and $y$ axes respectively. 
$U^T$ is the time-reversed sequence and  $\overline{U}$ stands 
for the sequence with inverted phases. }}
\end{table}

\section{Average Hamiltonian Theory \label{sec3}}

In this section we compare DD sequences 
constructed from time symmetric and asymmetric building blocks in the 
framework of average Hamiltonian theory \cite{Haeberlen1,ath}. 
While the zero-order average Hamiltonian of the asymmetric sequence is the same as that of the symmetric sequence,
this is no longer the case for the higher order terms.
In particular, if a cycle is symmetric, that is 
if $\tilde{H}(t) = \tilde{H}(\tau_c -t)$ for $0 \leq t \leq \tau_c$, where $\tilde{H}(t)$ represents the 
Hamiltonian in the toggling frame \cite{Haeberlen1,ath}, it can 
be shown  that all odd order terms
of the average Hamiltonian vanish \cite{burum}.
Clearly, this condition can only be fulfilled, if the timing of the sequence is symmetric,
as in the example of Fig.  \ref{fig1}a.

We first consider the 
sequence XY-4, which is our basic building block.
Our system consists of a single qubit, which we describe as a spin 1/2,
and an environment, which consists of a spin-bath.
The Hamiltonian describing the system plus environment  is  then
\begin{eqnarray}
H = H_S + H_{SE} + H_E,  
\label{HH}
\end{eqnarray}
where $H_S = \omega_S S_z $ is the system Hamiltonian,
$\mathbf{S} = (S_x,S_y,S_z)$ is the spin vector operator of the system qubit
 and $\omega_S$ is the Zeeman frequency of the system.
$H_E$ is the environment
Hamiltonian, which does not commute with $H_{SE}$ in general but is not specified further.
$H_{SE}$  is the system-environment interaction:
\begin{eqnarray}
H_{SE} &=&  \sum_k b_k S_z I_z^k .
\end{eqnarray}
In the following, we work in a resonant rotating frame, where  $\omega_S = 0$
and therefore $H_S = 0$. 
$\mathbf{I^k} = (I^k_x,I^k_y,I^k_z)$ is 
the spin vector operator 
of the $k^{th}$ environment spin,  
$b_k$  is the coupling constant between the system and the $k^{th}$ spin of the environment. 

In our case, the dominant source of experimental imperfections are flip-angle errors.
The actual pulse propagator for a nominal $\pi$ rotation around an axis defined by $\phi$ is then
\begin{eqnarray}
R(\phi) &=& e^{- i (1 + \epsilon) \pi S_{\phi}}
\label{rot}
\end{eqnarray}
where $\epsilon$ is the relative flip angle error, 
$S_{\phi} = cos \phi S_x + sin \phi S_y$, and $\phi$ is the phase of the pulse. 
We can write the zeroth ($\overline{H_0}$) and first ($\overline{H_1}$) order terms of the average 
Hamiltonian for the time symmetric XY-4 sequence
\begin{eqnarray}
\overline{H_0^{S}} &=& H_E \label{H0S} \\
\overline{H_1^{S}} &=& \frac{5 \epsilon^2 \pi^2}{16 \tau} S_z   -  \sum_k b_k { \frac{\epsilon \pi}{32} (S_x + S_y) I_z^k}.
\label{H1S}
\end{eqnarray} 
Details of the calculation are given in the appendix.
The zeroth order average Hamiltonian matches exactly the target Hamiltonian,
and for perfect pulses ($\epsilon=0$), the first order term vanishes, $\overline{H_1^{S}} = 0$, as expected for any symmetric sequence. 
For finite pulse errors, the first-order term contains a rotation of the spin qubit around the $z$ axis
by an angle $ 5 \epsilon^2 \pi^2/4$. 
This term results from the accumulated flip angle errors and is independent of the environment.
Since this term is proportional to the square of the flip angle error $\epsilon$,
it generates a rotation in the same direction for all spins, independent of the actual
field that they experience.

The second term in eq. (\ref{H1S}), in contrast, is linear in $\epsilon$. 
For an optimal setting of the pulse, $\epsilon$ is distributed symmetrically around zero
and the resulting evolution due to this term does not lead to an overall precession,
but to a loss of amplitude.
This term combines pulse errors and the system-bath interaction. 
It arises from the fact that pulses that do not 
implement a $\pi$ rotation cannot properly refocus the system-environment interaction. 

Now we compare 
these results with the average Hamiltonian of the time asymmetric form of XY-4:
\begin{eqnarray}
\overline{H_0^{A}} &=& H_E \label{H0A} \\
\overline{H_1^{A}} &=&  \frac{5 \epsilon^2 \pi^2}{16 \tau} S_z    - \sum_k b_k \frac{ \epsilon \pi }{16} S_x I_z^k  \nonumber \\
          & & + i \tau S_z \sum_k b_k [I_z^k, H_E],
\label{H1A}
\end{eqnarray}
The most striking difference from the symmetric case is the appearance of a new term  
which is a commutator between the internal Hamiltonian of the environment $H_E$ and the system-environment 
interaction Hamiltonian $H_{SE}$. 

Under ideal conditions, the first-order average Hamiltonian vanishes for the symmetric building block,
but not for the asymmetric case. 
For the asymmetric case the third term, which is proportional to $[H_{SE},H_E]$ remains.
The commutator describes the time dependence of the system-environment interaction
due to the environmental Hamiltonian $H_E$.
This difference from the symmetric case may be understood in terms of the different positions 
of the echoes shown in Fig. \ref{fig1} and Fig. \ref{cpmg1}:
In the asymmetric sequence, the time between echoes is twice as long as in the symmetric sequence,
which means that a time-dependent environment has a bigger effect.
In the symmetric sequence, the effect of the time-dependent environment appears only in the next-higher
order term. 

Rules for improving the performance of multiple pulse sequences were discussed, e.g., 
in the context of broadband heteronuclear decoupling \cite{levitt3}
or for the compensated Carr-Purcell sequences \cite{xy}. 
If we combine a XY-4 cycle with its time-reversed image to an XY-8 cycle,
we obtain a time-symmetric cycle, even if we start from the non-time symmetric block.
Nevertheless, we expect different results for the two cases.
An explicit calculation of the average Hamiltonian for the 
combined cycle shows that 
$\overline{H_0^{S}} = \overline{H_0^{A}} = H_E$ 
and  $\overline{H_1^{S}} = \overline{H_1^{A}} = 0$,
i.e. all deviations from the ideal Hamiltonian vanish to first order.
This remains true for finite pulse errors: the symmetry of the sequence cancels the effect
of pulse errors in all odd-order average Hamiltonian terms.

We therefore proceed to calculate the second order terms. 
For simplicity, we do not calculate the general expression, but consider two limiting cases.
First, we assume that the environmental Hamiltonian vanishes, $H_E = 0$.
The second order term then becomes
\begin{eqnarray}
\overline{H_2^{S}} &=&  \frac{13 \epsilon^3 \pi^3 }{1536 \tau}(S_x + S_y) + \sum_k \frac{\epsilon^2 \pi^2 b_k}{384} S_z I_z \label{H2Sa}\\  
\overline{H_2^{A}} &=&  \overline{H_2^{S}} + \sum_k \frac{\epsilon b_k^2 \tau}{368} S_y \label{H2Aa}.
\end{eqnarray}
Again, the average Hamiltonian for the sequence built from asymmetric blocks
contains an additional error term, which depends on the pulse error and the square
of the system-environment interaction.

As the second limiting case, we assume ideal pulses but non-vanishing environmental Hamiltonian, $H_E \neq 0$.
The second order terms then become
\begin{eqnarray}
\overline{H_2^{S}} &=&  \frac{\tau^2}{8} [[H_E,H_{SE}],H_E - \frac{1}{3} H_{SE}] \label{H2Sb}\\  
\overline{H_2^{A}} &=&  \overline{H_2^{S}} + \frac{\tau^2}{8} [[H_E,H_{SE}],7 H_E -H_{SE}]\label{H2Ab}.
\end{eqnarray}

As for the XY-4 sequence, the time-dependence of the environment, represented by the commutator 
$[H_E,H_{SE}]$ has the bigger effect if the sequence uses an asymmetric building block and therefore
generates echoes with bigger time delays between them.

\section{experimental Results \label{sec4}}

\subsection{Setup and system}

For the experimental tests we  used natural abundance $^{13}$C nuclear spins in the CH$_2$ groups
of a polycrystalline adamantane sample as the system qubit. 
The carbon spins are coupled to nearby $^1$H nuclear spins by heteronuclear magnetic dipole interaction
corresponding to $H_{SE}$.
The protons are coupled to each other by the homonuclear dipolar interaction,
which corresponds to $H_E$ and does not commute with $H_{SE}$.
The system environment interaction is therefore not static and the carbon spins
experience a fluctuating environment \cite{pra}. 
Under our conditions, the 
interaction between the carbon nuclei can be neglected and the
decoherence mechanism is a pure dephasing process \cite{pra}, the evolution 
of the system and environment is thus described by the Hamiltonian (\ref{HH}). 

The experiments were 
performed on a homebuilt 300 MHz
solid-state NMR spectrometer. The basic experimental scheme 
consisted of: a state preparation
period, during which we prepared the carbon spins in a superposition state
oriented along the $y$ direction, a variable evolution 
period, where DD sequences were applied, and the final read out period where we 
determine the final state by quantum state tomography \cite{chuang,oliveira}. 

\begin{figure}[htbp]
\vspace*{13pt}
\begin{center}
{\includegraphics[width=9.0cm]{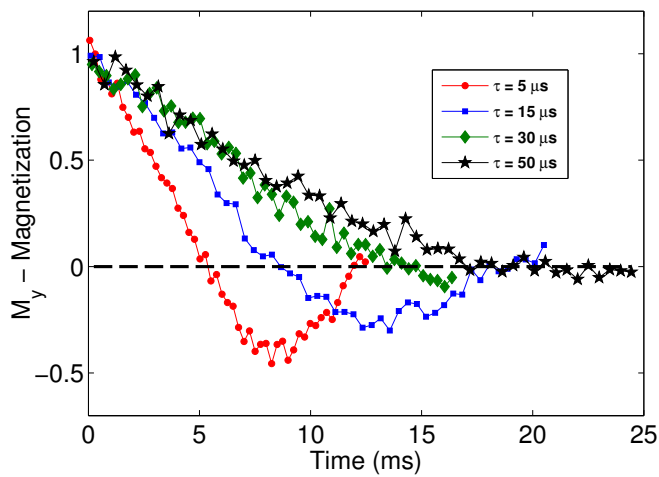}}
\end{center}
\vspace*{13pt}
\caption{\label{mag} Decay  of the $M_y$-magnetization for the symmetric version 
of XY-4 and different delays ($\tau$) between pulses. }
\end{figure}

Figure \ref{mag} shows the signal decay for the symmetric version 
of XY-4 and different delays $\tau$ between the refocusing pulses.
For the shortest cycle times, we observe shorter decays and oscillations.
As discussed in Ref. \cite{pra}, this is an indication that in this regime pulse
imperfections play the dominant role.
From the decay curves, we extract decay 
times as the times where the magnetization 
has decayed to $1/e$ of the initial value.

\subsection{Measured decay times}

\begin{figure}[htbp]
\vspace*{13pt}
\begin{center}
{\includegraphics[width=9.0cm]{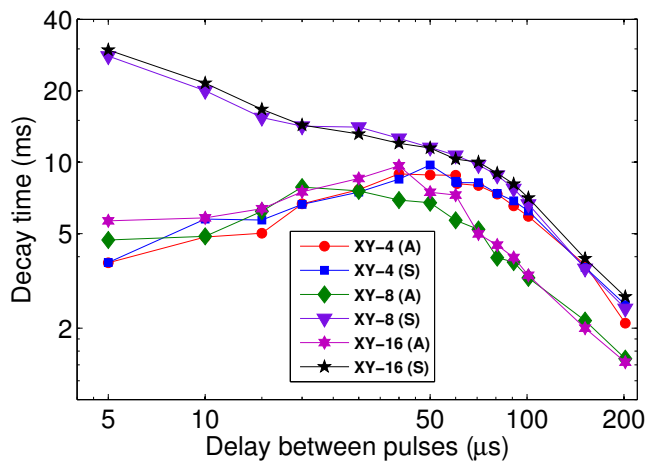}}
\end{center}
\vspace*{13pt}
\caption{\label{t2x} Decay time of the $M_y$-magnetization for different XY-$n$ sequences.  }
\end{figure}

Figure \ref{t2x} shows the decay times of the $M_y$-magnetization 
as a function of the delay $\tau$ between the pulses. 
For delays between 200 $\mu$s and 50 $\mu$s, the decoupling 
performance improves as the delays between the pulses are reduced. However, as the delay between 
the pulses becomes shorter than 50 $\mu$s, 
the decay time decreases again, in agreement with 
what was observed in \cite{pra,prl}: in this region, pulse errors become 
more important than the coupling to the environment.
This occurs equally for both, the symmetric 
and the asymmetric XY-4 sequence. 

If we concatenate the XY-4 sequence with its time-reversed version to the XY-8 sequence,
we obtain qualitatively different behavior for the two different versions of XY-4:
If we start from the symmetric form of XY-4, the resulting XY-8(S) sequence shows improved 
decoupling performance for increasing pulse rate, without saturating.
This is a clear indication that in this case, the concatenation eliminates the effect of pulse
imperfections and generates a robust, well-compensated sequence.
In strong contrast to this, concatenation of the asymmetric version of XY-4 to XY-8(A)
does not lead to a significant improvement: the decay times for XY-8(A) are identical to those
of the two XY-4 sequences, within experimental uncertainty.
A further concatenation to XY-16 does not change this behavior. 
The qualitatively different behavior of the sequences using symmetric versus asymmetric building blocks
clearly shows that for the asymmetric versions, the pulse errors dominate, while the symmetric ones
compensate for the effect of pulse errors.

\subsection{Tomographic analysis}

\begin{figure}[htbp]
\vspace*{13pt}
\begin{center}
\subfigure{\includegraphics[width=8.0cm]{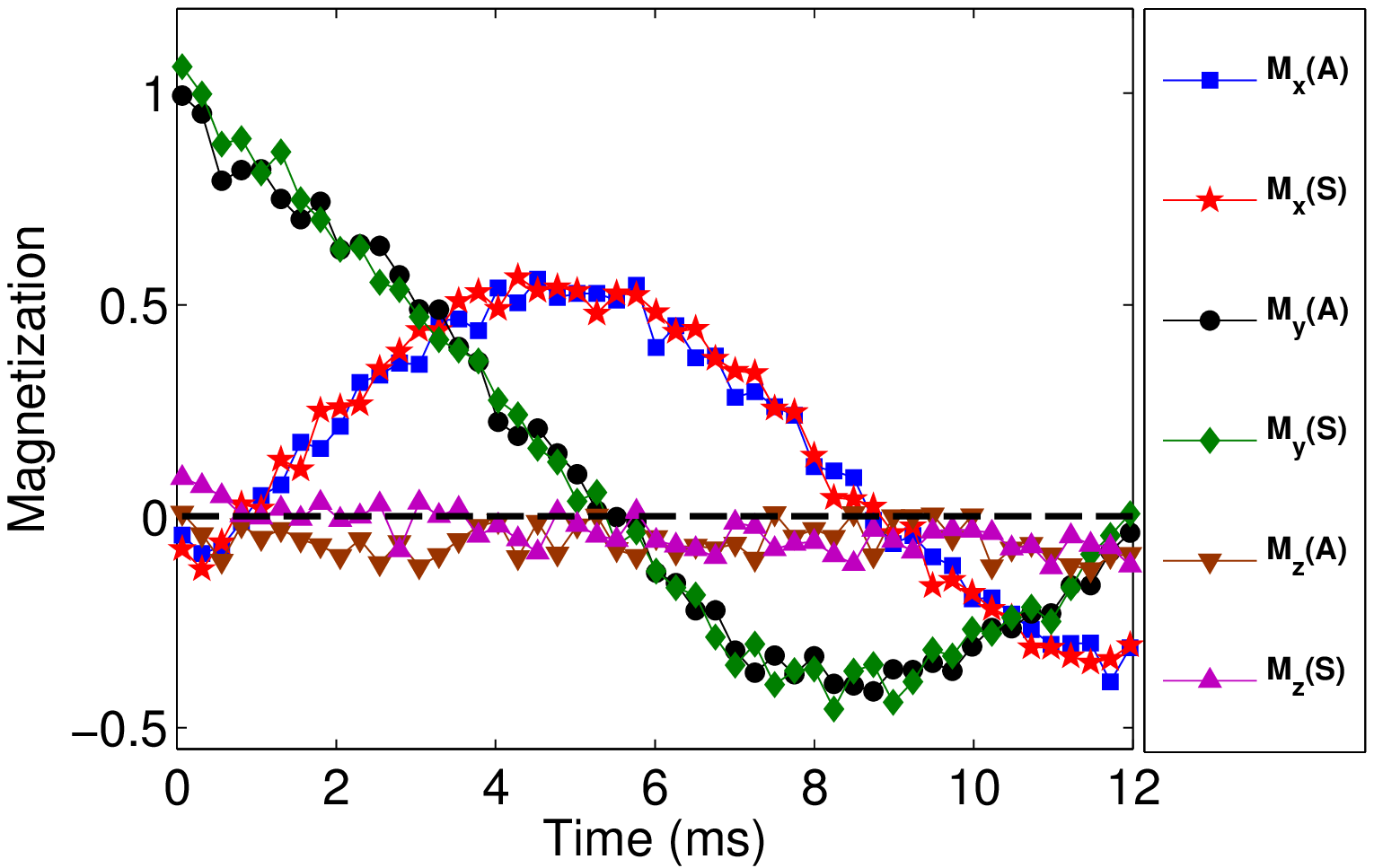}} \\
\subfigure{\includegraphics[width=10.0cm]{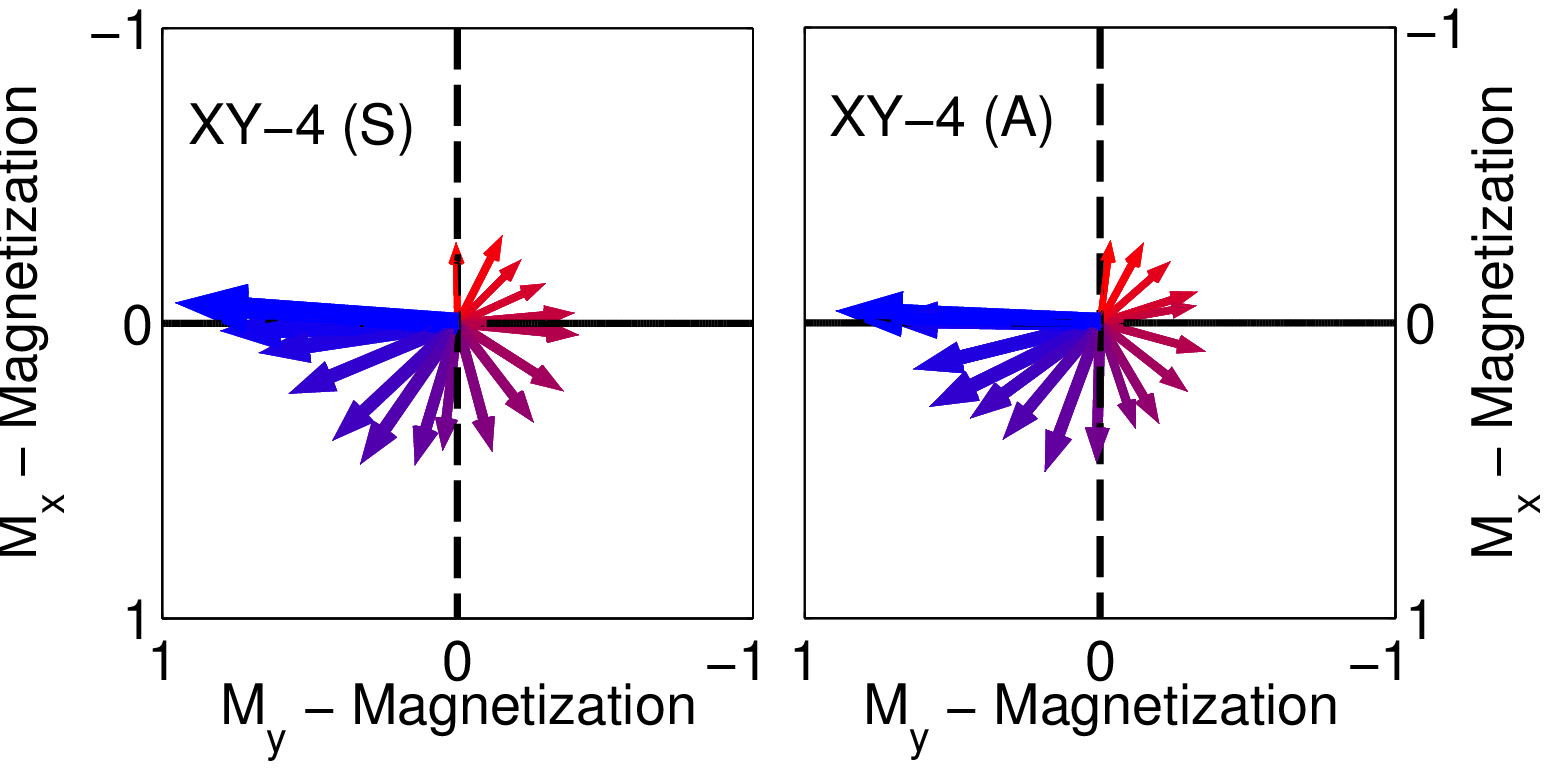}}
\end{center}
\vspace*{13pt}
\caption{\label{ppxy4} Evolution of the magnetization during the
symmetric (A) and asymmetric (S) versions of the 
XY-4 sequence for pulse spacings of $\tau = 10 \mu s$. 
The top panel shows the evolution of the 
magnetization components as a function of time. 
The bottom panel represents the
Bloch vector in the $xy$ plane at different times. The color 
code in the lower panel denotes the time evolution, blue for the 
initial state and red for the final state.}
\end{figure}

For a more detailed picture of the process that reduces the signal for high pulse rates,
we applied state tomography of the evolving qubit by measuring all three components
along the $x$, $y$ and $z$-direction.
Figure \ref{ppxy4} shows the observed data for both versions of the XY-4 sequence.
The oscillation of the $x$- and $y$ components and the constantly small value of the
$z$-component are a clear indication of a precession around the $z$-axis,
in addition to the loss of signal amplitude.
This combination of precession and reduction of amplitude is also shown in the lower part of 
Fig. \ref{ppxy4}, where the arrows show the xy-components of the magnetization for different
times during the sequence. 
According to eqs (\ref{H1S}) and (\ref{H1A}), the precession around the $z$-axis 
originates from the pulse error term $5 \epsilon^2 \pi^2/(16 \tau) S_z$, which is 
proportional to $\epsilon^2$ and is the same
for the symmetric and the asymmetric sequence, in excellent agreement with the observed behavior.

\begin{figure}[htbp]
\vspace*{13pt}
\begin{center}
\subfigure{\includegraphics[width=8.0cm]{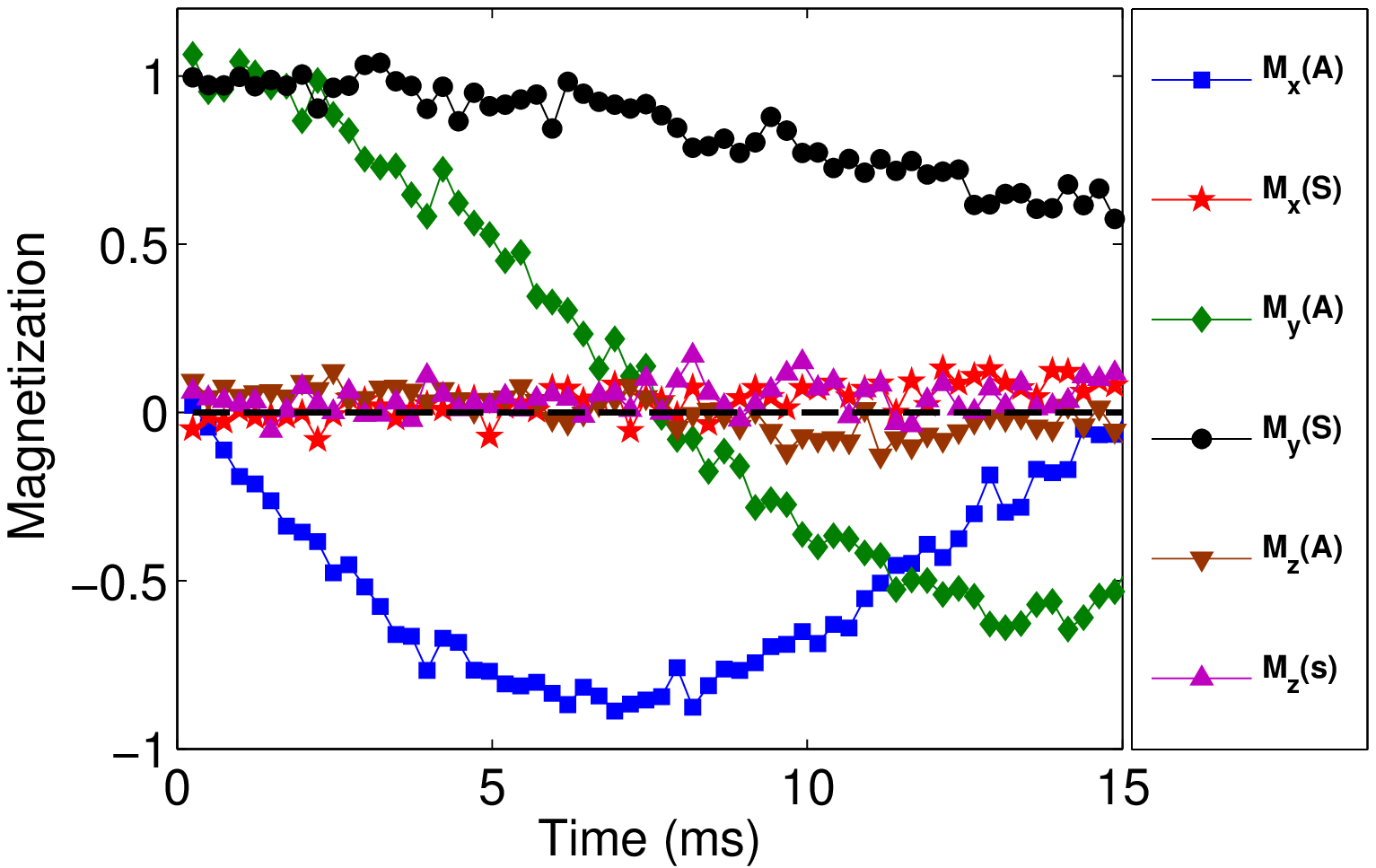}} \\
\subfigure{\includegraphics[width=10.0cm]{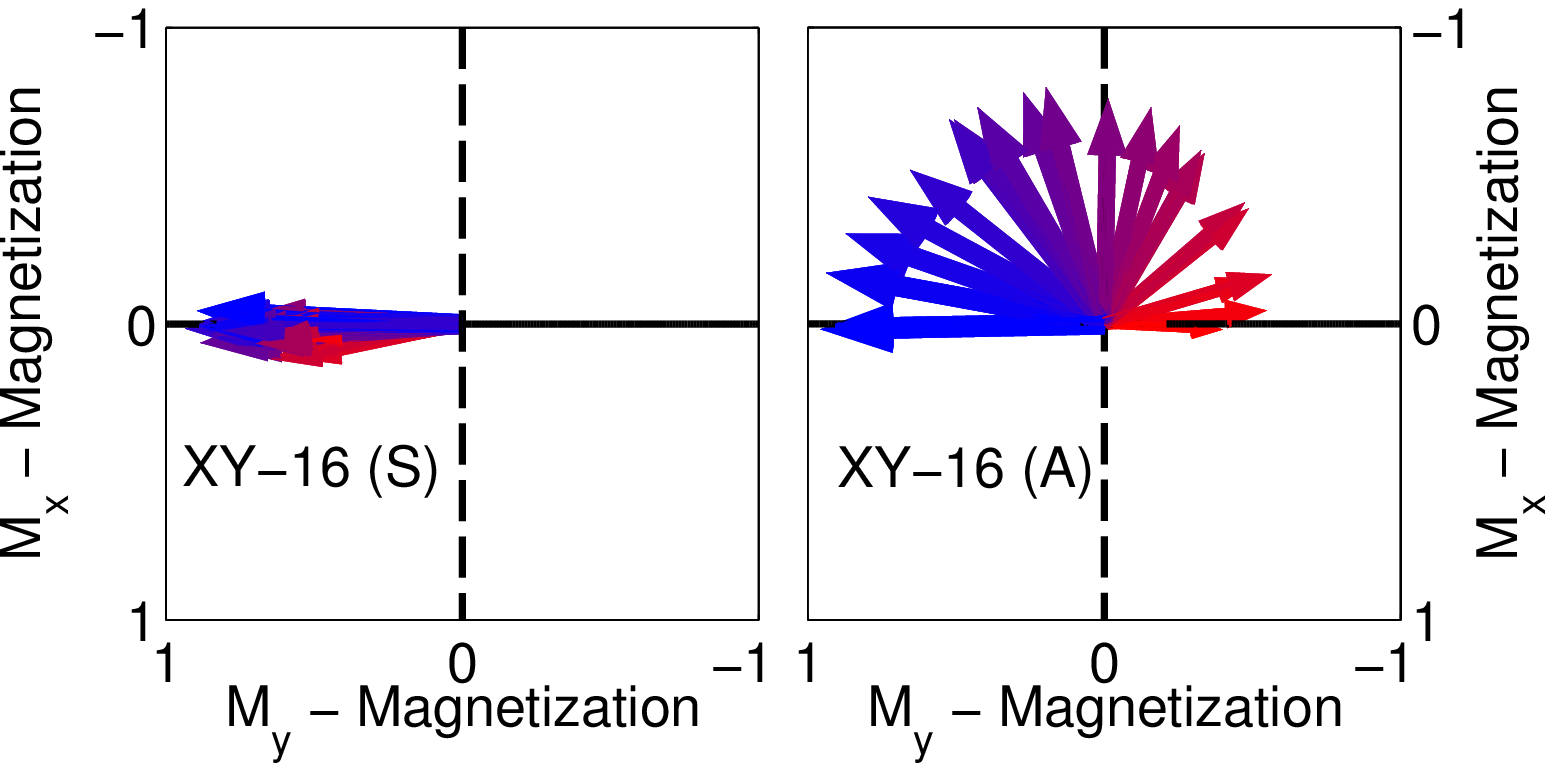}}
\end{center}
\vspace*{13pt}
\caption{\label{ppxy8}  
Evolution of the magnetization during the
symmetric (A) and asymmetric (S) versions of the 
XY-16 sequence for pulse spacings of $\tau = 10 \mu s$. 
The top panel shows the evolution of the 
magnetization components as a function of time. 
The bottom panel represents the
Bloch vector in the $xy$ plane at different times. The color 
code is the same as in Fig. \ref{ppxy4}.}
\end{figure}

Figure \ref{ppxy8} shows the corresponding data for the two XY-16 sequences.
Here, as well as in the case of XY-8 (data not shown), we also observe a precession for the XY-16(A) sequence, but for the
sequence with symmetric building blocks, the oscillation is not observed.
Again, these results indicate that the sequences built from symmetric XY-4 blocks
have smaller average Hamiltonians and therefore show better performance than those built from asymmetric blocks.

\begin{figure}[htbp]
\vspace*{13pt}
\begin{center}
{\includegraphics[width=9.0cm]{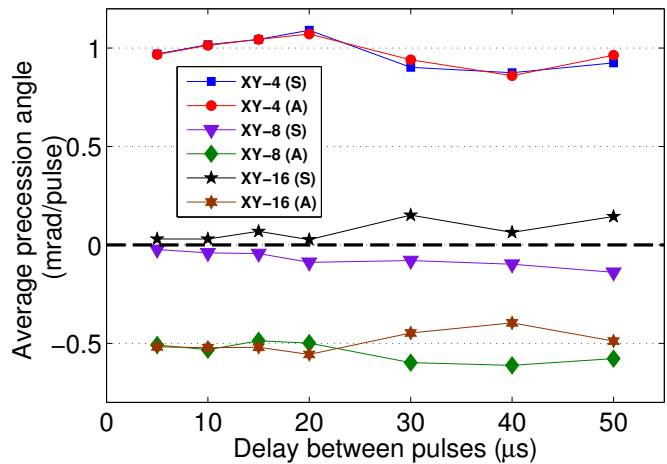}}
\end{center}
\vspace*{13pt}
\caption{\label{pxy} Average precession angle of the transverse magnetization during the different XY-$n$ sequences.  }
\end{figure}

If we change the spacing between the pulses, the behavior remains the same.
In Fig.  \ref{pxy}, we show the measured precession angle around the $z$-axis divided by the number of pulses.
The precession is indistinguishable from zero for the compensated XY-8(S) and XY-16(S) sequences.
For other sequences, it is significant and independent 
of the delay between the pulses.

\begin{figure}[htbp]
\vspace*{13pt}
\begin{center}
{\includegraphics[width=9.0cm]{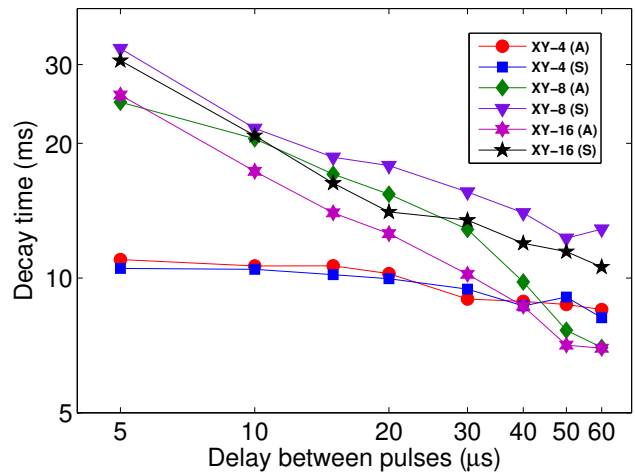}}
\end{center}
\vspace*{13pt}
\caption{\label{t2n} Decay time of the total magnetization for XY-$n$ sequences.  }
\end{figure}

The precession of the magnetization around the $z$-axis that we observe for some of the sequences
causes a deviation of the system from the desired evolution and reduces therefore the fidelity of the process.
However, compared to a dephasing process, it is easier to correct and can in principle be compensated if it is known.
We therefore compared not only the reduction of the magnetization amplitude along the initial direction,
but also the total magnetization left in the system, which eliminates the effect of the precession.
Figure \ref{t2n} shows the decay times of the total magnetization for different XY-$n$ sequences. 
For short delays between the pulses, the difference between sequences built by  
symmetric and asymmetric building blocks is small,
indicating that the main difference is related to the precession originating from the 
pulse errors, which is better
compensated by concatenating symmetric building blocks.
For  pulse delays longer than $\tau \approx 15 \mu$s, we start to see again that the 
symmetric versions of XY-8(S) and XY-16(S) are superior to the asymmetric versions.
At this point, the time dependence of the environment plays a bigger role and reduces
the efficiency of the refocusing \cite{pra,prl}.
In agreement with eqs (\ref{H2Aa}) and  (\ref{H2Ab}), these effects are bigger for those sequences 
that use asymmetric building blocks. 
 
\section{Discussion and Conclusions \label{sec5}}

Dynamical decoupling is becoming a standard technique for extending the lifetime of 
quantum mechanical coherence.
Many different sequences have been put forward for reducing the effect of the environmental noise
on the system.
Since the number of possible sequences is infinite, a relatively straightforward approach for
designing improved sequences consists in concatenating different building blocks in such a way
that the resulting cycle has a smaller overall average Hamiltonian than that of its component blocks.
In this paper we consider the XY-4 sequence as a basic building block.
Since different versions of the XY-4 sequence were proposed in the literature,
some of them symmetric in time, others asymmetric, we compare these two
versions and in particular the different sequences that result when they are 
concatenated with time-inverted and phase-shifted copies.
Since time-symmetric sequences generate average Hamiltonians in which all odd-order
terms vanish for ideal pulses,
it is expected that they perform better than non-symmetric but otherwise
identical sequences.
Experimentally, we could not verify this for the XY-4 sequence, since pulse errors dominate
the behavior under our experimental conditions.
However, in the case of the CPMG sequence, where pulse errors are insignificant, 
we could clearly verify this expectation.

The symmetry of the basic building blocks is also important when they are concatenated
to higher order sequences, such as the XY-8 and XY-16 sequences \cite{xy}.
In this case, the odd order terms vanish in the average Hamiltonians of both sequences,
but the second order terms of the sequences that are built from asymmetric blocks
contain additional unwanted terms. 
The experimental data are in agreement with this observation:
sequences consisting of time symmetric building blocks  perform significantly  better 
than the corresponding sequences formed by time asymmetric blocks. 

In order to understand the decay processes during the DD sequences, we performed quantum state
tomography as a function of time. 
The results from these measurements show two different contributions to the overall fidelity loss:
A precession around the $z$-axis, which we could attribute to the combined effect of flip-angle errors
and an overall reduction of the amplitude, which results from the system-environment interaction.
For short delays between the pulses and correspondingly large number of pulses,
the pulse error term is the dominating effect. 
Again, the symmetric and asymmetric version of the XY-4 sequence show similar performance.
However, as we use them as building blocks of the higher-order XY-8 and XY-16 sequences,
we find that the effect of the pulse errors is almost perfectly compensated if we use the symmetric
building blocks, while a significant effect remains when asymmetric blocks are concatenated.

\begin{figure}[htbp]
\vspace*{13pt}
\begin{center}
{\includegraphics[width=9.0cm]{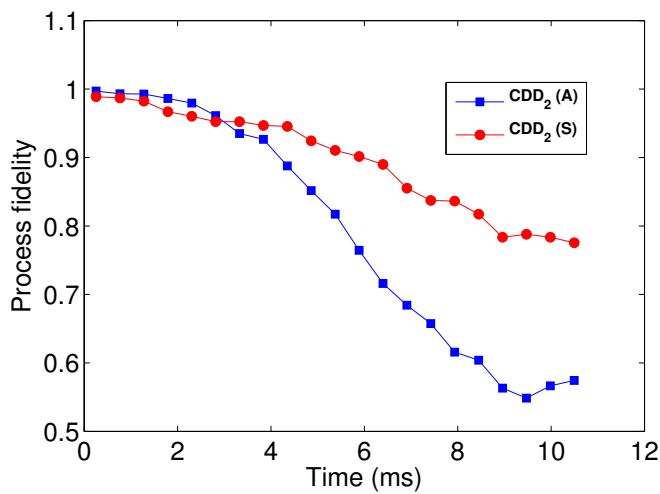}}
\end{center}
\vspace*{13pt}
\caption{\label{fcdd}  Process fidelities of the symmetric vs. asymmetric version 
of CDD-2 as a function of time for an 
average delay $\tau = 12.5 \mu\mathrm{s}$ between the refocusing pulses. }
\end{figure}

While we have analyzed the effect of symmetry  mostly for the XY-$n$ sequences,
this can clearly be generalized.
As we showed in Fig. \ref{cpmg}, the symmetric version of the CPMG sequence
shows significantly better decoupling performance than the asymmetric version.
The same concept can also be applied to the CDD sequences, which are generated by
inserting XY-4 sequences inside the delays of a lower-order CDD sequence \cite{cdd}.
The conventional concatenation scheme  \cite{cdd} uses asymmetric building blocks.
Here, we used the symmetric XY-4 sequence as the building-blocks, 
and we modified the concatenation scheme in such a way that the symmetry is preserved
and the delays between the pulses are identical at all levels of concatenation.
The conventional (asymmetric) version CDD$_n$(A) is iterated as $[CDD_{n-1}-X-CDD_{n-1}-Y]^2$.
In contrast to that, we construct the symmetrized version CDD$_n$(S) as 
$[\sqrt{CDD_{n-1}}-X-CDD_{n-1}-Y-\sqrt{CDD_{n-1}}]^2$ \cite{prl}. 
For $n=1$, we have $CDD_1$ = XY-4.
In Fig. \ref{fcdd} we compare the process fidelities \cite{pfidel,nielsen} for the 
two versions of the CDD-2 sequence.
Clearly, the symmetrized version CDD$_2$(S) shows a significantly 
improved performance, compared to the standard,
asymmetric version CDD$_2$(A).  

The results presented in this paper show 
clearly that it is time reversal symmetry 
is a useful tool for improving dynamic decoupling sequences.
The symmetric sequences often perform better and never worse than non-symmetric sequences,
at no additional cost.

\ack
We acknowledge useful discussions with Daniel Lidar.
This work is supported by the DFG through Su 192/24-1.

\appendix
\section{Average Hamiltonian Calculation}
In this Appendix we show how the average Hamiltonian 
was calculated. The essence of average-Hamiltonian theory is
that a cyclic evolution, $U(t)$, can be described 
by an effective evolution governed by a time-independent 
Hamiltonian $\overline{H}$. When $U$ is comprised  by a sequence of 
unitary operations, i.e.  $U = e^{O_n} \cdots e^{O_2}e^{O_1}$, the average 
Hamiltonian can be approximately computed by recursive applications 
of the Baker-Campbell-Hausdorff formula

\begin{eqnarray}
 log(e^{A}e^{B}) &\approx& A + B + \frac{1}{2}[A,B] + \nonumber \\
&& \frac{1}{12} \left( [A,[A,B]] + [[A,B],B] \right)
\label{bch}
\end{eqnarray}

To approximate the average Hamiltonian of the XY-4 sequence, consider the following sequence:
\begin{eqnarray}
[\tau_i - R_1 - \tau - R_2 - \tau - R_3 - \tau - R_4 -\tau_f],
\label{seq}
\end{eqnarray} 
where $R_1 = R_3 = R(X)$, $R_2 = R_4 = R(Y)$ and $R({\phi})$ is a the pulse propagator 
defined in Eq. \ref{rot}. 
In the  asymmetric form 
of XY-4, $\tau_i = \tau$ and $\tau_f = 0$. For $\tau_i = \tau_f = \tau/2$, we obtain the symmetric form of XY-4. 
The total sequence propagator is
\begin{eqnarray}
 U &=& e^{-i H \tau_f} \left(\prod_{k=2}^{4} R_k e^{-i H \tau} \right) R_1 e^{-i H \tau_i} , 
\label{propag1}
\end{eqnarray}
 where $H$ is the Hamiltonian of Eq. (\ref{HH}).  

To account for flip angle errors, we decompose the propagator into a product of the ideal pulse propagator sandwiched 
between two additional evolutions: 
\begin{eqnarray}
 R_{\phi} &=& e^{-i (1 + \epsilon) \pi S_{\phi}} \nonumber \\
          &=& e^{-i H_{\phi} \frac{t_p}{2}} e^{-i \pi S_{\phi}}  e^{-i H_{\phi} \frac{t_p}{2}} 
\label{rotdec}
\end{eqnarray} 
where $ H_{\phi} =  \frac{\epsilon \pi}{t_p} S_{\phi}$ and $t_p$ is pulse length. Substituting  
equation (\ref{rotdec}) in (\ref{propag1}) and using the following approximation:
\begin{eqnarray}
 e^{-i H_{\phi} \frac{t_p}{2}}  e^{-i H \tau} &\approx& e^{-i (H \tau + H_{\phi} \frac{t_p}{2})} \nonumber \\
                                        &\approx& e^{-i (H  + \frac{1}{2 \tau} \epsilon \pi S_{\phi}) \tau} \nonumber  \\
                                        &\approx& e^{-i H^{'} \tau },
\end{eqnarray} 
the new sequence propagator is then rewritten as
\begin{eqnarray}
U &=& e^{-i H_5^{'} \tau_f} \left(\prod_{k=2}^{4} R_k e^{-i H_k^{'}  \tau} \right) R_1 e^{-i H_1^{'}  \tau_i} ,
\label{propag1b}
\end{eqnarray}
where 

\begin{eqnarray}
\tau_i H_1^{'}  &=&\tau_i  H  + \frac{t_p}{2} H_{X}   \\
\tau H_{k=2,3,4}^{'}  &=&\tau H  + \frac{t_p}{2} \left( H_{X} + H_{Y} \right)   \\
\tau_f H_5^{'}  &=&\tau_i  H  + \frac{t_p}{2} H_{Y}.
\end{eqnarray} 
The calculation can be simplified by transforming the Hamiltonians to a new frame 
after each pulse, the so-called toggling frame. The Hamiltonians $\widetilde{H_k}$  in this new frame 
are given by:

\begin{eqnarray}
 \tau_i \widetilde{H_1} &=& \tau_i(H_E + H_{SE}) + \frac{t_p}{2} H_X \label{tgh1}\\ 
 \tau \widetilde{H}_{k=2,3,4} &=& \tau [H_E + (-1)^{k+1} H_{SE}]  \nonumber \\
&&  - t_p [(2 \delta_{k,2} + 1) H_X + (2 \delta_{k,4} +1) H_Y] \label{tgh2}\\
 \tau_f \widetilde{H_5} &=& \tau_f(H_E + H_{SE}) + \frac{t_p}{2} H_Y \label{tgh3} 
\end{eqnarray}

and the  sequence propagator is:
\begin{eqnarray}
 U \approx  \prod_{k=1}^{5} e^{-i \widetilde{H_k} \tau_k} 
\label{propag1c}
\end{eqnarray}

The final step consists of the recursive applications of eq. (\ref{bch}) to 
eq. (\ref{propag1c}). An explicit calculation of the zeroth and first 
order terms leads to the 
equations (\ref{H0S}) and (\ref{H1S}) for the symmetric case and (\ref{H0A}) and (\ref{H1A}) for 
the asymmetric case.  The calculation for XY-8 follows the same procedure 
as described for XY-4. Here the sequence is 
comprised by eight pulses and nine delays (see table \ref{tab1}), this leads to the total 
propagator analog to \ref{propag1b}:   
\begin{eqnarray}
U &=& e^{-i H_9^{'} \tau_f} \left(\prod_{k=2}^{8} R_k e^{-i H_k^{'}  \tau} \right) R_1 e^{-i H_1^{'}  \tau_i} ,
\label{propag2b}
\end{eqnarray}
Transforming the Hamiltonians $H_k^{'}$ to the toggling frame, one can 
show that $\widetilde{H_k} = \widetilde{H}_{1-k}$, the Hamiltonians 
$\widetilde{H}_{k=1,2,3,4}$ are the same  as in Eqs. (\ref{tgh1}) - Eqs. (\ref{tgh2})  and  
$\tau \widetilde{H_5}  = \tau (H_E + H_{SE})  + t_p H_{Y}$.


\section*{References}
\begin{thebibliography}{50}

\bibitem{viola}
L.~Viola, E.~Knill, and S.~Lloyd.
\newblock Dynamical decoupling of open quantum systems.
\newblock {\em Phys. Rev. Lett.}, 82:2417, 1999.

\bibitem{yang}
Wen Yang, Zhen-Yu Wang, and Ren-Bao Liu.
\newblock Preserving qubit coherence by dynamical decoupling.
\newblock {\em Front. Phys.}, 6:1, 2010.

\bibitem{ddgate1}
K.~Khodjasteh and L.~Viola.
\newblock Dynamically error-corrected gates for universal quantum computation.
\newblock {\em Phys. Rev. Lett.}, 102:080501, Feb 2009.

\bibitem{ddgate2}
K.~Khodjasteh, D.~A. Lidar, and L.~Viola.
\newblock Arbitrarily accurate dynamical control in open quantum systems.
\newblock {\em Phys. Rev. Lett.}, 104:090501, Mar 2010.

\bibitem{ddgate3}
J.~R. West, D.A. Lidar, B.~H. Fong, and M.~F. Gyure.
\newblock High fidelity quantum gates via dynamical decoupling.
\newblock {\em Phys. Rev. Lett.}, 105:230503, Dec 2010.

\bibitem{ddgate4}
H~.K. Ng, D.~A. Lidar, and J.~Preskill.
\newblock Combining dynamical decoupling with fault-tolerant quantum
  computation.
\newblock {\em Phys. Rev. A}, 84:012305, Jul 2011.

\bibitem{dqc1dd}
Sergio Boixo and Rolando~D. Somma.
\newblock Parameter estimation with mixed-state quantum computation.
\newblock {\em Phys. Rev. A}, 77:052320, May 2008.

\bibitem{noise1}
J.~Bylander, S.~Gustavsson, F.~Yan, F.~Yoshihara, K.~Harrabi, G.~Fitch, D.~G.
  Cory, Y.~Nakamura, J-S Tsai, and W.~D. Oliver.
\newblock Noise spectroscopy through dynamical decoupling with a
  superconducting flux qubit.
\newblock {\em Nat. Phys.}, 7:565, 2011.

\bibitem{noise2}
I.~Almog, Y.~Sagi, G.~Gordon, G.~Bensky, G.~Kurizki, and N.~Davidson.
\newblock Direct measurement of the system-environment coupling as a tool for
  understanding decoherence and dynamical decoupling.
\newblock {\em J. Phys. B: At. Mol. Opt. Phys.}, 44(15):154006, 2011.

\bibitem{noise3}
G.~A. Alvarez and D.~Suter.
\newblock Dynamical decoupling noise spectroscopy.
\newblock arXiv:1106.3463 (2011) (Submitted).

\bibitem{magnetometer1}
J.~M. Taylor, P.~Cappellaro, L.~Childress, L.~Jiang, D.~Budker, P.~R. Hemmer,
  A.~Yacoby, R.~Walsworth, and M.~D. Lukin.
\newblock High-sensitivity diamond magnetometer with nanoscale resolution.
\newblock {\em Nat. Phys.}, 4:810, 2008.

\bibitem{meriles}
C.~A. Meriles, L.~Jiang, G.~Goldstein, J.~S. Hodges, J.~Maze, M.~D. Lukin, and
  P.~Cappellaro.
\newblock {\em J. Chem. Phys.}, 133:124105, 2010.

\bibitem{magnetometer2}
L.~T. Hall, C.~D. Hill, J.~H. Cole, and L.~C.~L. Hollenberg.
\newblock Ultrasensitive diamond magnetometry using optimal dynamic decoupling.
\newblock {\em Phys. Rev. B}, 82:045208, Jul 2010.

\bibitem{magnetometer3}
G.~de~Lange, D.~Rist\`e, V.~V. Dobrovitski, and R.~Hanson.
\newblock Single-spin magnetometry with multipulse sensing sequences.
\newblock {\em Phys. Rev. Lett.}, 106:080802, Feb 2011.

\bibitem{exp4}
J.~Du, X.~Rong, N.~Zhao, Y.~Wang, J.~Yang, and R.~B. Liu.
\newblock Preserving electron spin coherence in solids by optimal dynamical
  decoupling.
\newblock {\em Nature}, 421:1265, 2009.

\bibitem{exp3}
M.~J. Biercuk, H.~Uys, A.~P. VanDevender, N.~Shiga, W.~M. Itano, and J.~J.
  Bollinger.
\newblock Optimized dynamical decoupling in a model quantum memory.
\newblock {\em Nature}, 458:996, 2009.

\bibitem{warren}
E.~R. Jenista, A.~M. Stokes, R.~T. Branca, and W.~S. Warren.
\newblock Optimized, unequal pulse spacing in multiple echo sequences improves
  refocusing in magnetic resonance.
\newblock {\em J. Chem. Phys.}, 131:204510, 2009.

\bibitem{pra}
G.~A. {\'A}lvarez, A.~Ajoy, X.~Peng, and D.~Suter.
\newblock Performance comparison of dynamical decoupling sequences for a qubit
  in a rapidly fluctuating spin bath.
\newblock {\em Phys. Rev. A}, 82:042306, 2010.

\bibitem{exp1}
G.~deLange, Z.~H. Wang, D.~Rist{\'e}, V.~V. Dobrovitski, and R.~Hanson.
\newblock Universal dynamical decoupling of a single solid-state spin from a
  spin bath.
\newblock {\em Science}, 330:60, 2010.

\bibitem{exp2}
C.~A. Ryan, J.~S. Hodges, and D.~G. Cory.
\newblock Robust decoupling techniques to extend quantum coherence in diamond.
\newblock {\em Phys. Rev. Lett.}, 105:200402, 2010.

\bibitem{ashok}
Ashok Ajoy, Gonzalo~A. \'Alvarez, and Dieter Suter.
\newblock Optimal pulse spacing for dynamical decoupling in the presence of a
  purely dephasing spin bath.
\newblock {\em Phys. Rev. A}, 83(3):032303, Mar 2011.

\bibitem{prl}
A.~M. Souza, G.~A. \'Alvarez, and D.~Suter.
\newblock Robust dynamical decoupling for quantum computing and quantum memory.
\newblock {\em Phys. Rev. Lett.}, 106:240501, 2011.

\bibitem{jpb}
Z-H. Wang and V.~Dobrovitski.
\newblock Aperiodic dynamical decoupling sequences in the presence of pulse
  errors.
\newblock {\em J. Phys. B: At. Mol. Opt. Phys.}, 44:154004, 2011.

\bibitem{Haeberlen1}
U.~Haeberlen and J.~S. Waugh.
\newblock Coherent averaging effects in magnetic resonance.
\newblock {\em Phys. Rev.}, 175:453--467, Nov 1968.

\bibitem{mansfield}
P.~Mansfield.
\newblock Symmetrized pulse sequences in high resolution nmr in solids.
\newblock {\em J. Phys. C: Solid State Phys.}, 4(11):1444, 1971.

\bibitem{rhim1}
W.~K. Rhim, D.~D. Elleman, and R.~W. Vaughan.
\newblock Analysis of multiple pulse nmr in solids.
\newblock {\em J. Chem. Phys.}, 59:3740, 1973.

\bibitem{rhim2}
D.~P. Burum and W.~K. Rhim.
\newblock Analysis of multiple pulse nmr in solids ii.
\newblock {\em J. Chem. Phys.}, 71:944, 1979.

\bibitem{levitt1}
M.~H. Levitt.
\newblock {\em J. Chem. Phys.}, 128:052205, 2008.

\bibitem{levitt2}
M.~H. Levitt.
\newblock Symmetry-based pulse sequences in magic-angle spinning solid-state
  nmr.
\newblock In Encyclopedia of NMR (Wiley, 2002).

\bibitem{burum}
D.~P. Burum.
\newblock Magnus expansion generator.
\newblock {\em Phys. Rev. B}, 24:3684, 1981.

\bibitem{pines1}
A.~J. Shaka and A.~Pines.
\newblock {\em J. Magn. Reson.}, 71:495, 1987.

\bibitem{morris}
J.~T. Ngo and P.~G. Morris.
\newblock {\em J. Magn. Reson.}, 74:122, 1987.

\bibitem{maudsley}
A.A Maudsley.
\newblock Modified carr-purcell-meiboom-gill sequence for nmr fourier imaging
  applications.
\newblock {\em J. Magn. Reson.}, 69(3):488, 1986.

\bibitem{ath}
R.~R. Ernst, G.~Bodenhausen, and A.~Wokaum.
\newblock {\em Principles of nuclear magnetic resonance in one and two
  dimensions}.
\newblock Clarendon Press, United Kingdom, 1987.

\bibitem{chuang}
I.~L. Chuang, N.~Gershenfeld, M.~G. Kubinec, and D.~W. Leung.
\newblock Bulk quantum computation with nuclear magnetic resonance: theory and
  experiment.
\newblock {\em Proc. R. Soc. Lond. A}, 454:447, 1998.

\bibitem{oliveira}
I.~S. Oliveira, R.~S. Sarthour, E.~R. deAzevedo, T.~J. Bonagamba, and J.~C.~C.
  Freitas.
\newblock {\em NMR Quantum information processing}.
\newblock Elsevier, Netherland, 2007.

\bibitem{cp}
H.~Y. Carr and E.~M. Purcell.
\newblock Effects of diffusion on free precession in nuclear magnetic resonance
  experiments.
\newblock {\em Phys. Rev.}, 94:630--638, May 1954.

\bibitem{mg}
S.~Meiboom and D.~Gill.
\newblock Modified spin echo method for measuring nuclear relaxation times.
\newblock {\em Rev. Sci. Instrum.}, 29:688, 1958.

\bibitem{cdd}
K.~Khodjasteh and D.~A. Lidar.
\newblock Fault-tolerant quantum dynamical decoupling.
\newblock {\em Phys. Rev. Lett.}, 95:180501, 2005.

\bibitem{xy}
T.~Gullion, D.~B. Baker, and M.~S. Conradi.
\newblock New compensated carr-purcell sequences.
\newblock {\em J. Magn. Reson.}, 89:479, 1990.

\bibitem{Khodjasteh}
K.~Khodjasteh and D.~A. Lidar.
\newblock Performance of deterministic dynamical decoupling schemes:
  Concatenated and periodic pulse sequences.
\newblock {\em Phys. Rev. A}, 75:062310, 2007.

\bibitem{sdd}
L.~F. Santos and L.~Viola.
\newblock {\em New J. Phys.}, 10:083009, 2008.

\bibitem{preskill}
H.~K. Ng, D.~A. Lidar, and J.~Preskill.
\newblock Combining dynamical decoupling with fault-tolerant quantum
  computation.
\newblock {\em Phys. Rev. A}, 84:012305, Jul 2011.

\bibitem{levitt3}
M.~H. Levitt, R.~Freeman, and T.~Frenkiel.
\newblock {\em J. Magn. Reson.}, 50:157, 1982.

\bibitem{pfidel}
I.~L. Chuang and M.~A. Nielsen.
\newblock Prescription for experimental determination of the dynamics of a
  quantum black box.
\newblock {\em J. Mod. Opt.}, 44:2455, 1997.

\bibitem{nielsen}
M.~A. Nielsen and I.~L. Chuang.
\newblock {\em Quantum Computation and Quantum Information}.
\newblock Cambridge University Press, Cambridge, 2000.

\end{thebibliography}
\end{document}